\documentclass[12pt,epsf]{article}

\usepackage{graphicx}
\usepackage{mathrsfs}
\usepackage{amssymb}
\usepackage{amsmath}

\newcommand{\adsiv}{${\rm AdS}_4/{\rm CFT}_3$}

\newcommand{\adsv}{${\rm AdS}_5/{\rm CFT}_4$}

\newcommand{\be}{\begin{equation}}

\newcommand{\ee}{\end{equation}}

\newcommand{\ba}{\begin{array}}

\newcommand{\ea}{\end{array}}

\begin{document}
\begin{titlepage}
\vspace{.5in}
\begin{flushright}
CQUeST-2010-0370
\end{flushright}
\vspace{0.5cm}

\begin{center}
{\Large\bf Quantum finite-size effects for dyonic magnons in the $AdS_{4}\times\mathbb{CP}^{3}$}\\
\vspace{.4in}

{Changrim Ahn$^{\S}$, Minkyoo Kim$^{\dag}$, and Bum-Hoon Lee$^{\dag}$}\footnote{E-mail:
ahn@ewha.ac.kr, mkim80@sogang.ac.kr, bhl@sogang.ac.kr}
\vskip 1cm
{\small $^\S$ \it Department of Physics and Institute for the Early Universe\\
Ewha Womans University, Seoul 120-750, S. Korea }\\
\vskip .5cm
{\small $^\dag$ \it Department of Physics and Center for Quantum Spacetime\\
 Sogang University, Seoul 121-742, S. Korea}\\

\vspace{.5in}
\end{center}
\begin{center}
{\large\bf Abstract}
\end{center}
\begin{center}
\begin{minipage}{4.75in}

We compute quantum corrections to finite-size effects for various dyonic giant magnons in the
$AdS_4\times\mathbb{CP}^3$ in two different approaches.
The off-shell algebraic curve method is used to quantize the classical string configurations
in semi-classical way and to compute the corrections to the string energies.
These results are compared with the $F$-term L\"uscher formula based on the $S$-matrix
of the ${\rm AdS}_4/{\rm CFT}_3$.
The fact that the two results match exactly provides another stringent test for the all-loop integrability
conjecture and the exact $S$-matrix based on it.

%\pacs{Valid PACS appear here}% PACS, the Physics and Astronomy
                             % Classification Scheme.
%\Keywords{suggested Keywords}%Use showkeys class option if keyword
                             % display desired

\end{minipage}
\end{center}
\end{titlepage}

\newpage

\section{Introduction\label{sec1}}
The AdS/CFT correspondence \cite{AdS/CFT}
between string theories on ${\rm AdS}$ backgrounds and
supersymmetric conformal gauge theories has been studied very actively and
produced many exciting developments in understanding non-perturbative structures
of the string and gauge theories.
In the ${\rm AdS}_4/{\rm CFT}_3$ duality the three-dimensional conformal field theory
is ${\cal N}=6$ super Chern-Simons (CS) theory
with $SU(N)\times SU(N)$ gauge symmetry and level $k$.  This model,
which was first proposed by Aharony, Bergman, Jafferis, and Maldacena (ABJM)
\cite{Aharony:2008ug}, is believed to be dual to M-theory on $AdS_4\times
S^7/Z_k$.
Furthermore, in the planar limit of $N, k\to\infty$ with a
fixed value of 't Hooft coupling $\lambda=N/k$, the ${\cal N}=6$ CS is
believed to be dual to type IIA superstring theory on $AdS_4\times
\mathbb{CP}^3$.

In the strong coupling limit $\lambda>>1$, the string theory on $AdS_4\times\mathbb{CP}^3$ is classically
integrable \cite{Arutyunov:2008if,Stefanski:2008ik,Gromov:2008bz} and probably maintains the
integrability for any value of $\lambda$ which would lead to the all-loop Bethe ansatz equations (BAEs)
as conjectured in \cite{Gromov:2008qe}.
%%%%%%%%%%%%%%%%%%%%
A discrepancy in one-loop correction to the energy of a folded spinning string from
the all-loop Bethe ansatz results \cite{MR, AAB, Kr} can be resolved by a non-zero one-loop correction
in the central interpolating function $h(\lambda)$ \cite{MRT}.
%%%%%%%%%%%%%%%%%%%%%
Based on the assumption that the ABJM model in the planar limit is quantum integrable along with
centrally extended $su(2|2)$ symmetry, the exact $S$-matrix has been
conjectured and used to confirm the above exact BAEs in \cite{Ahn:2008aa}.
However, this working hypothesis of all-loop integrability needs to be checked in all available methods.
An efficient way is to compute and compare finite-size effects from both $S$-matrix and string theory sides.
A most successful way for $S$-matrix side is the L\"uscher formula which computes
small energy shift of on-shell particle states due to the finite-size of the system.

In the classical string theory side, various classical string configurations which correspond to the on-shell
particle states have been identified.
In addition to such classical string configurations as small giant magnon (GM) ($\mathbb{CP}^{1}$) \cite{Gaiotto:2008cg}
and small dyonic GM ($\mathbb{CP}^{2}$) \cite{Abbott:2009um,Hollowood:2009sc} which have also found in ${\rm AdS}_5/{\rm CFT}_4$,
new type of classical string solutions like pair of small GM ($\mathbb{RP}^{2}$) \cite{Gaiotto:2008cg,Grignani:2008is,Grignani:2008te},
pair of small dyonic GM ($\mathbb{RP}^{3}$) \cite{Ahn:2008hj}, and big GM
\cite{Shenderovich:2008bs,Hollowood:2009tw,Kalousios:2009mp,Suzuki:2009sc}
have been also discovered in $AdS_4\times\mathbb{CP}^3$.
The finite-size effects for these string configurations have been computed by either solving the superstring
sigma model directly such as Neumann-Rosochatius reduction or the algebraic curve method
\cite{Ahn:2008hj,Shenderovich:2008bs,Lukowski:2008eq,Abbott:2009um}.
Leading finite-size corrections for the ${\rm AdS}_4/{\rm CFT}_3$ string configurations have been
confirmed by the $\mu$-term L\"uscher formula for non-dyonic ($\mathbb{CP}^{1}$ and $\mathbb{RP}^{2}$) GM \cite{Bombardelli:2008qd} and dyonic ($\mathbb{RP}^3$) GM \cite{Ahn:2008wd}.

The purpose of this paper is to provide some more stringent tests of the exact $S$-matrix in the strong coupling limit.
In the string theory side, we implement the off-shell quantization of the algebraic curve method developed in
\cite{Gromov:2008ec} for the various string configurations.
These are compared with the $F$-term  L\"uscher formula in the $S$-matrix side.
These two independent computations are shown to be identical, which provides a stringent test for both
$S$-matrix and all-loop integrability conjecture.
This paper is organized as follows.
We will present in sect.2 the quantum corrections for various classical string configurations
using the off-shell algebraic curve formulation.
The $S$-matrix and the $F$-term L\"uscher formula based on it are used in sect.3 to derive the finite-size
effects. We will also generalize the one loop energy shifts to general multi-magnons in sec.4.
We conclude the paper with a few remarks in sect.5.

\section{Algebraic curve method\label{sec2}}

\subsection{GMs in the algebraic curve}
In quantum field theory, quantum effects can be obtained by considering fluctuations of fields
in an effective Lagrangian.
However, this method can be very complicated for the string theory on nontrivial background.
A more efficient way to compute semi-classical effects of string solutions has been developed
in \cite{Gromov:2007aq,Gromov:2008ie,Gromov:2008ec} on the basis of classical integrability of
the string sigma-model on $AdS$ backgrounds.
The classical integrability of type IIA superstring on $AdS_{4}\times \mathbb{CP}^{3}$ has been shown
in \cite{Arutyunov:2008if,Stefanski:2008ik} and the related algebraic curve has been constructed
in \cite{Gromov:2008bz}.

In the algebraic curve framework, different string solutions are mapped to different sets of
eigenvalues of the classical transfer matrix.
These are realized through a set of quasi-momenta $q_{i}(x)$ $i=1,\cdots,10$ depending on a spectral parameter $x$, which
are defined on a multi-sheet Riemann surface with particular analytic properties.
In particular, solutions living mostly in $\mathbb{CP}^{3}$ have the following quasi-momenta as proposed
in \cite{Gromov:2008bz, Gromov:2008qe, Lukowski:2008eq}:
\begin{eqnarray}
q_{1}&=& -q_{10}= \frac{\alpha x}{x^{2}-1} \nonumber\\
q_{2}&=& -q_{9}= \frac{\alpha x}{x^{2}-1} \nonumber\\
q_{3}&=& -q_{8}= \frac{\alpha x}{x^{2}-1}+ G_{u}\left(0\right)
-G_{u}\left(\frac{1}{x}\right) + G_{v}\left(0\right)
-G_{v}\left(\frac{1}{x}\right) + G_{r}\left(x\right)
-G_{r}\left(0\right)+G_{r}\left(\frac{1}{x}\right) \nonumber\\
q_{4}&=& -q_{7}= \frac{\alpha x}{x^{2}-1}+ G_{u}\left(x\right)
+ G_{v}\left(x\right) - G_{r}\left(x\right)+G_{r}\left(0\right)-G_{r}\left(\frac{1}{x}\right)\nonumber\\
q_{5}&=& -q_{6}= G_{u}\left(x\right)-G_{u}\left(0\right)
+G_{u}\left(\frac{1}{x}\right) - G_{v}\left(x\right)+G_{v}\left(0\right)-G_{v}\left(\frac{1}{x}\right).\nonumber
\end{eqnarray}
Here $\alpha$ is related to the energy $\Delta$ by $\alpha=\frac{\Delta}{2g}$.
This ansatz satisfies several analytic properties \cite{Gromov:2008bz}.

At large $x$, quasi-momenta are related with physical conserved quantities:
\begin{equation}
\lim_{x\rightarrow\infty}\left(\begin{array}{c}
q_{1}(x)\\
q_{2}(x)\\
q_{3}(x)\\
q_{4}(x)\\
q_{5}(x)\end{array}\right)\simeq\frac{1}{2gx}\left(\begin{array}{c}
\Delta\\
\Delta\\
J_{1}\\
J_{2}\\
J_{3}
\end{array}\right).\label{asymp}
\end{equation}
$J$'s are angular momenta of strings moving in $\mathbb{CP}^{3}$.
Because we are interested in string solutions moving mostly in $\mathbb{CP}^{3}$,
we have set spins in $AdS$ to zero.
These quasi-momenta are related with each other by inversion symmetry:
\begin{equation}
\left(\begin{array}{c}
q_{1}\left(1/x\right)\\
q_{2}\left(1/x\right)\\
q_{3}\left(1/x\right)\\
q_{4}\left(1/x\right)\\
q_{5}\left(1/x\right)\end{array}\right)=\left(\begin{array}{c}
0\\
0\\
\pi m\\
\pi m\\
0\end{array}\right)+\left(\begin{array}{c}
-q_{2}(x)\\
-q_{1}(x)\\
-q_{4}(x)\\
-q_{3}(x)\\
+q_{5}(x)\end{array}\right).\label{inv}
\end{equation}

Virasoro constraints in string $\sigma$-model manifest in the algebraic curve through the synchronization
of poles at $x=\pm1$ \cite{Gromov:2008bz}.
\begin{equation}
\lim_{x\rightarrow\pm1}\left(\begin{array}{c}
q_{1}(x)\\
q_{2}(x)\\
q_{3}(x)\\
q_{4}(x)\\
q_{5}(x)\end{array}\right)\simeq\frac{1}{2(x\mp1)}\left(\begin{array}{c}
\alpha_{\pm}\\
\alpha_{\pm}\\
\alpha_{\pm}\\
\alpha_{\pm}\\
0\end{array}\right).\label{vira}
\end{equation}

In contrast with the $AdS_{5} \times S^{5}$ case, we need three resolvents $G_{u},G_{v}$ and $G_{r}$.
Subscripts $u$,$v$,$r$ correspond to different kinds of excitations,
which are also related with Dynkin labels of $su(4)$.
One can reproduce the charges of different string solutions by choosing different resolvents.
The resolvents for the GM solutions, in particular, are log-cut distributions of Bethe roots
and can be thought of as degeneration of two cuts \cite{Minahan:2006bd,Minahan:2008re,Sax:2008in}.
At infinite $J$ the GM resolvents are defined as \cite{Minahan:2006bd}
\begin{equation}
G_{\rm magnon}=-i \log\left(\frac{x-X^{+}}{x-X^{-}}\right),\nonumber
\end{equation}
while at finite size we have to think of the GM as a degenerate case of a two-cut solution with a resolvent
\cite{Lukowski:2008eq,Minahan:2008re,Sax:2008in}
\begin{equation}
G_{\rm finite}=-2i \log\left(\frac{\sqrt{x-X^{+}}+\sqrt{x-Y^{+}}}{\sqrt{x-X^{-}}+\sqrt{x-Y^{-}}}\right).\nonumber
\end{equation}

There are three kinds of GM solutions in $\mathbb{CP}^{3}$
\cite{Gaiotto:2008cg,Grignani:2008is,Shenderovich:2008bs,Hollowood:2009tw,Kalousios:2009mp,Suzuki:2009sc}.
If $G_{u} = G_{\rm magnon}$ and $G_{v}=G_{r}=0$, then we obtain the small GM solution on $\mathbb{CP}^{2}$.
If we consider the $v$-type resolvent, we obtain small magnon with a reversed sign in the charge $Q$.
If we set $G_{u}=G_{v}=G_{\rm magnon}$ and $G_{r}=0$,
then we obtain a pair of small GMs, also called a $\mathbb{RP}^{3}$ solution.
This solution corresponds to a GM excitation in each $su(2)$ sector of the gauge theory,
and in particular we will consider only configurations with the same momentum in each sector.
Finally, the big GM solution corresponds to $G_{u}=G_{v}=G_{r}=G_{\rm magnon}$, which is
two-parameter but one-charge solution because $Q$ is zero \cite{Abbott:2009um}.

Finite-size corrections of GMs in the algebraic curve can be computed from fluctuations of
quasi-momenta.
When we compute classical finite-size corrections, we use $G_{\rm finite}$ and consider
log-cut distribution of Bethe roots with small square-root cut at the tips.
As deviation from usual log-cut is very small, we could take the leading term of the
fluctuation of quasi-momenta \cite{Minahan:2008re,Sax:2008in}.
On the other hand, if we are interested in quantum corrections,
we have to add extra poles
to quasi-momenta with $G_{\rm magnon}$ \cite{Gromov:2007aq}.
Such extra poles give fluctuations of energy by summing over on-shell frequencies.
As we will explain in detail in sect.2.2, on-shell frequencies are efficiently obtained
from the off-shell frequencies.
We first compute the fluctuation frequencies by using the off-shell method \cite{Gromov:2008ec}
in section 2.2 then evaluate energy shift in section 2.3.

\subsection{Off-shell frequency of Magnons}

One-loop computation using the algebraic curve method has been developed in \cite{Gromov:2007aq,Gromov:2008ie}.
As explained above, fluctuations of superstring fields are mapped onto quasi-momenta  fluctuations.
These correspond to adding extra poles connecting sheets of the Riemann surface on which
quasi-momenta are defined \cite{Gromov:2007aq}.
From the quasi-momenta perturbations, we can compute fluctuation frequencies.

There are two ways to compute fluctuation frequencies. In the on-shell method, we compute  fluctuation
frequencies by adding extra small poles to all physical polarization pairs. The position of these extra poles is determined by $q_{i}\left(x^{ij}_{n}\right)-q_{j}\left(x^{ij}_{n}\right)=2\pi n_{ij}$.
But this method involves repeated lengthy calculations for general solution, such as generic two-cut ones.
%As authors of  show us, we can do more clever. We'll use such more efficient method, off-shell method. In this wa, we only consider to adding particular extra poles corresponding to polarization we want to consider. For instance, if we just think about $\left\left(1,5\right)$ polarization then we only add extra poles in $q_{1}$ and $q_{5}$. And in this computation, $\delta\Delta$ that can be read from large x asymptotic is just $\Omega_{15}$.
More efficient way is the off-shell method where we need to consider only additional poles of quasi-momenta
for particular polarization pair \cite{Gromov:2008ec}. In this way, we don't fix the position of extra pole which we consider.
The general ansatz for fluctuations of quasi-momenta can be written using analytic properties of algebraic curve.
Then, we evaluate fluctuations of quasi-momenta at large $x$ to determine the physical conserved charges.
We can easily obtain off-shell fluctuation frequencies from these equations.

We will use this efficient off-shell method to compute one-loop effects for dyonic GMs.
%In fact we check that off-shell frequency in these two ways are the same each other.
Fluctuation frequencies $\Omega_{ij}$ of all three types of GMs of \adsiv\,\ are closely related with those of \adsv\,\
with one distinctive difference that there are two kinds of physical modes in \adsiv\,\ - heavy and light modes.
Heavy modes, however, do not contribute to one-loop leading term since it is suppressed exponentially.

Because functional form of off-shell frequencies of all three GMs are identical,
we only show off-shell frequencies for the small magnon.
A small magnon is obtained by $G_{u}=G_{\rm magnon}$ and $G_{v}=G_{r}=0$ \cite{Lukowski:2008eq,Abbott:2009um}.
The quasi-momenta of small magnon are given by
\begin{eqnarray}
q_{1}&=& -q_{10}= \frac{\alpha x}{x^{2}-1} \nonumber\\
q_{2}&=& -q_{9}= \frac{\alpha x}{x^{2}-1} \nonumber\\
q_{3}&=& -q_{8}= \frac{\alpha x}{x^{2}-1}-i \log\left(\frac{X^{+}}{X^{-}}\right)
+i \log\left(\frac{\frac{1}{x}-X^{+}}{\frac{1}{x}-X^{-}}\right)+\tau \nonumber\\
q_{4}&=& -q_{7}= \frac{\alpha x}{x^{2}-1}-i \log\left(\frac{x-X^{+}}{x-X^{-}}\right)+\tau\nonumber\\
q_{5}&=& -q_{6}= -i \log\left(\frac{x-X^{+}}{x-X^{-}}\right)+i \log\left(\frac{X^{+}}{X^{-}}\right)
-i \log\left(\frac{\frac{1}{x}-X^{+}}{\frac{1}{x}-X^{-}}\right).\label{ssmall}
\end{eqnarray}

Here, $\tau$ is a twist which is introduced to make single magnon satisfy the usual periodic boundary condition.
In the case of small magnon, $\tau= -\frac{p}{2}$.
(The twist for both a pair of small magnons and a big magnon is $-p$.)

On-shell fluctuation energy is defined as follows:
\begin{equation}
\Omega^{ij}_{n}= -\kappa_{ij}\delta_{i,1}+2g \lim_{x\rightarrow \infty} x \delta_{n}^{ij}q_{1}\left(x\right).\nonumber
\end{equation}
Here, $\kappa_{ij}=2$  for $\left(i,j\right)=\left(1,10\right),\left(2,9\right)$ and $\kappa_{ij}=1$ for other pairs.
Also, $n_{ij}$ are  mode numbers and moduli of the algebraic curve.
 They satisfy that $q_{i}^{+}-q_{j}^{-}=2\pi n_{ij}$.
To obtain off-shell energy from on-shell energy, we just change $n$-dependence in the above expression to
quasi-momentum dependence as follows \cite{Gromov:2008ec}:
\begin{equation}
\Omega^{ij}\left(y\right)= \Omega^{ij}_{n}|_{n\rightarrow\frac{q_{i}\left(y\right) - q_{j}\left(y\right)}{2\pi}}.\nonumber
\end{equation}
There are 16 polarization modes for type IIA superstring in $AdS_{4}\times \mathbb{CP}^{3}$.
They consist of 8 light modes which are $(i,5)$ or $(i,6)$ pairs and 8 heavy modes for other pairs.
We have to compute all types of fluctuation frequencies for these polarization modes.
But, as explained first in \cite{Gromov:2008ec} and studied in \cite{Bandres:2009kw} for \adsiv,
off-shell frequencies are related to each other by the inversion symmetry.
The light modes are related by
\begin{eqnarray}
\Omega_{i5}\left(x\right)&=&\Omega_{i6}\left(x\right)\cr
\Omega_{25}\left(x\right)&=&\Omega_{15}\left(0\right)-\Omega_{15}\left(\frac{1}{x}\right)\cr
\Omega_{35}\left(x\right)&=&\Omega_{45}\left(0\right)-\Omega_{45}\left(\frac{1}{x}\right),\nonumber
\end{eqnarray}
and heavy modes by
\begin{eqnarray}
\Omega_{17}\left(x\right)&=&\Omega_{15}\left(x\right)+\Omega_{57}\left(x\right)=\Omega_{15}\left(x\right)
+\Omega_{45}\left(x\right)\cr
\Omega_{18}\left(x\right)&=&\Omega_{15}\left(x\right)+\Omega_{58}\left(x\right)=\Omega_{15}\left(x\right)
+\Omega_{35}\left(x\right)\cr
\Omega_{19}\left(x\right)&=&\Omega_{15}\left(x\right)+\Omega_{59}\left(x\right)=\Omega_{15}\left(x\right)
+\Omega_{25}\left(x\right)\cr
\Omega_{110}\left(x\right)&=&\Omega_{15}\left(x\right)+\Omega_{15}\left(x\right)=2\Omega_{15}\left(x\right)\cr
\Omega_{27}\left(x\right)&=&\Omega_{25}\left(x\right)+\Omega_{57}\left(x\right)=\Omega_{25}\left(x\right)
+\Omega_{45}\left(x\right)\cr
\Omega_{28}\left(x\right)&=&\Omega_{25}\left(x\right)+\Omega_{58}\left(x\right)=\Omega_{25}\left(x\right)
+\Omega_{35}\left(x\right)\cr
\Omega_{29}\left(x\right)&=&\Omega_{25}\left(x\right)+\Omega_{59}\left(x\right)=2\Omega_{25}\left(x\right)\cr
\Omega_{37}\left(x\right)&=&\Omega_{35}\left(x\right)+\Omega_{57}\left(x\right)=\Omega_{35}\left(x\right)
+\Omega_{45}\left(x\right).\nonumber
\end{eqnarray}
All $\Omega_{ij}$'s can be written in terms of only $\Omega_{15}$ and $\Omega_{45}$.

Now we calculate the off-shell frequencies by considering fluctuations of quasi-momenta in the form of extra poles \cite{Gromov:2008ec,Gromov:2007aq,Gromov:2008ie}.
The quasi-momenta fluctuations can be determined by properties (\ref{inv}) and (\ref{vira}).
For (1,5) polarization, we must add extra poles to $q_{1}$ and $q_{5}$.
If we write $\delta q_{1}$, $\delta q_{4}$, $\delta q_{5}$ to satisfy the analytic properties,
then $\delta q_{2}$ and $\delta q_{3}$ can be automatically written by inversion symmetry.
Note that $\delta q_{5}$ has self-inversion symmetry.
The ansatz is then written as follows:
\begin{eqnarray}
\delta q_{1}&=& \frac{A_{+}}{x+1} + \frac{A_{-}}{x-1} + \frac{\alpha\left(y\right)}{x-y}\cr
\delta q_{2}&=& -\frac{A_{+}}{\frac{1}{x}+1} - \frac{A_{-}}{\frac{1}{x}-1}
- \frac{\alpha\left(y\right)}{\frac{1}{x}-y}\cr
\delta q_{3}&=& -\frac{A_{+}}{\frac{1}{x}+1} - \frac{A_{-}}{\frac{1}{x}-1}
- \frac{B^{+}}{\frac{1}{x}-X^{+}} - \frac{B^{-}}{\frac{1}{x}-X^{-}}\cr
\delta q_{4}&=& \frac{A_{+}}{x+1} + \frac{A_{-}}{x-1} + \frac{B^{+}}{x-X^{+}} + \frac{B^{-}}{x-X^{-}}\cr
\delta q_{5}&=& -\frac{\alpha\left(y\right)}{x-y}-\frac{\alpha\left(y\right)}{\frac{1}{x}-y}
-\frac{\alpha\left(y\right)}{y}+\frac{B^{+}}{x-X^{+}} + \frac{B^{-}}{x-X^{-}}\cr
&&+\frac{B^{+}}{\frac{1}{x}-X^{+}} + \frac{B^{-}}{\frac{1}{x}-X^{-}}+\frac{B^{+}}{X^{+}}+\frac{B^{-}}{X^{-}},\nonumber
\end{eqnarray}
where $\alpha\left(x\right)=\frac{1}{2g} \frac{x^{2}}{x^{2}-1}$ .
The unknown constants $A_{\pm}$ and $B^{\pm}$ in the above ansatz are determined by the following equations
coming from the asymptotic behaviors of the fluctuations in (\ref{asymp}):
\begin{eqnarray}
A_{+}-A_{-}&=& \frac{\alpha\left(y\right)}{y}\cr
\frac{B^{+}}{X^{+}}+\frac{B^{-}}{X^{-}}&=&\frac{\alpha\left(y\right)}{y}\cr
A_{+}+A_{-}+\frac{\alpha\left(y\right)}{y^{2}}&=& \frac{\delta\Delta}{2g}\cr
A_{+}+A_{-}+\frac{B^{+}}{X^{+ 2}}+\frac{B^{-}}{X^{- 2}}&=&0\cr
A_{+}+A_{-}+B^{+}+B^{-}&=&0.\label{off}
\end{eqnarray}
In the above equations, we used the $N_{ij}$ in \cite{Shenderovich:2008bs}.
The fluctuation energy $\delta\Delta$ is given by the frequencies $\Omega_{ij}$
\begin{equation}
\delta\Delta=\sum_{ij,n} N_{ij}^{n}\Omega_{ij}^{n}.\label{del}
\end{equation}
For $(1,5)$ polarization, all $N_{ij}$ are zero except $N_{15}$. Also, we add just one pole ($N_{15}=1$).
Then, from (\ref{del}), we have $\delta\Delta=\Omega_{15}\left(y\right)$.
Now, we can solve $\delta\Delta$ from (\ref{off}),
\begin{equation}
\delta\Delta = \frac{1}{y^{2}-1} \left(1-y \frac{X^{+}+X^{-}}{X^{+}X^{-}+1}\right).\nonumber
\end{equation}

With the similar argument, we can also compute $\Omega_{45}\left(y\right)$
which turns out to be the same as $\Omega_{15}(y)$.
Using $\Omega_{15}\left(y\right)$ and $\Omega_{45}\left(y\right)$,
we can obtain frequencies
\begin{equation}
\Omega^{\rm light}_{ij}\left(y\right)=\frac{1}{y^{2}-1} \left(1-y \frac{X^{+}+X^{-}}{X^{+}X^{-}+1}\right),\label{lt}
\end{equation}
for the light modes, and
\begin{equation}
\Omega^{\rm heavy}_{ij}\left(y\right)=\frac{2}{y^{2}-1} \left(1-y \frac{X^{+}+X^{-}}{X^{+}X^{-}+1}\right),\label{hv}
\end{equation}
for the heavy modes, which are twice of those of light modes.

These are exactly the same results as those of \cite{Shenderovich:2008bs}.
Similarly, off-shell frequencies for other cases (pair of small and big magnon) can be evaluated.
In other cases, we also consider the most general ansatz for fluctuations of quasi-momenta and compare their asymptotic with the conserved charges.
Then, we get a set of similar equations as (\ref{off}). The result for other magnons is exactly same than that of the small magnon.
We can use the off-shell frequencies of the small magnon (\ref{lt}) and (\ref{hv}) for all other magnons.

\subsection{One-loop shifts of dyonic GMs}
The leading part of one-loop energy shift is given by the sum of fluctuation frequencies.
\begin{eqnarray}
\delta\Delta_{\rm one-loop}&=& \frac{1}{2}\sum_{ij}\sum_{n}\left(-1\right)^{F_{ij}}\Omega_{ij}^{n}
=\int \frac{dx}{2\pi i} \partial_{x}\Omega\left(x\right) \sum_{ij} \gamma_{ij}\left(-1\right)^{F_{ij}}
e^{-i\left(q_{i}-q_{j}\right)}.\nonumber
\end{eqnarray}
Here, $\gamma_{ij}= 1$ for light modes and $\gamma_{ij}= 2$ for heavy modes and $\Omega\left(x\right)$ in the last expression is off-shell energy. When we change from infinite summation over $n$ to integration over $x$, there are some non-trivial steps which we have to be careful \cite{Shenderovich:2008bs,Gromov:2008ie,Abbott:2010yb}.%We will not explain these because they are sufficiently explained \\

When we evaluate the above integral by using saddle-point approximation,
heavy modes can be suppressed because of the factor 2 in exponent.
This is related to the fact there are no $\frac{\alpha x}{x^{2}-1}$ term in $q_{5}$ or $q_{6}$.
Hence, only light modes are important in this computation.
This fact is consistent with the observation in \cite{Zarembo:2009au}
that only 8 light degrees of freedom are physical.
The other 8 heavy degrees of freedom of superstring theory are unstable.

So we need to compute $\sum_{ij}\left(-1\right)^{F_{ij}} e^{-i\left(q_{i}-q_{j}\right)}$
where the sum over $\left(i,j\right)$ pairs includes only the light modes.
Using quasi-momenta of the small magnon in (\ref{ssmall}), we obtain
\begin{eqnarray}
&&\sum_{ij} \left(-1\right)^{F_{ij}} e^{-i\left(q_{i}-q_{j}\right)}=
e^{\frac{-i\alpha x}{x^{2}-1}}\left[2\frac{\frac{1}{x}-X^{+}}{\frac{1}{x}-X^{-}}\frac{x-X^{+}}{x-X^{-}}
\frac{X^{+}}{X^{-}}+2\frac{\frac{1}{x}-X^{-}}{\frac{1}{x}-X^{+}}\frac{x-X^{-}}{x-X^{+}}\frac{X^{-}}{X^{+}}
-\frac{\frac{1}{x}-X^{+}}{\frac{1}{x}-X^{-}}\sqrt{\frac{X^{-}}{X^{+}}}\right.\nonumber\\
&&\left.-\frac{x-X^{-}}{x-X^{+}}\sqrt{\frac{X^{+}}{X^{-}}}
-\left(\frac{\frac{1}{x}-X^{+}}{\frac{1}{x}-X^{-}}\right)^{2}\frac{x-X^{+}}{x-X^{-}}
\left(\frac{X^{-}}{X^{+}}\right)^{\frac{3}{2}}-\frac{\frac{1}{x}-X^{-}}{\frac{1}{x}-X^{+}}
\left(\frac{x-X^{+}}{x-X^{-}}\right)^{2}\left(\frac{X^{+}}{X^{-}}\right)^{\frac{3}{2}}\right].\label{small}
\end{eqnarray}\\
The pair of small magnon is obtained by the following ansatz:
\begin{eqnarray}
q_{1}&=& -q_{10}= \frac{\alpha x}{x^{2}-1} \nonumber\\
q_{2}&=& -q_{9}= \frac{\alpha x}{x^{2}-1} \nonumber\\
q_{3}&=& -q_{8}= \frac{\alpha x}{x^{2}-1}-2i \log\left(\frac{X^{+}}{X^{-}}\right)+2i \log\left(\frac{\frac{1}{x}-X^{+}}{\frac{1}{x}-X^{-}}\right)-p \nonumber\\
q_{4}&=& -q_{7}= \frac{\alpha x}{x^{2}-1}-2i \log\left(\frac{x-X^{+}}{x-X^{-}}\right)-p\nonumber\\
q_{5}&=& -q_{6}= 0,\nonumber
\end{eqnarray}
which leads to
\begin{equation}
\sum_{ij} \left(-1\right)^{F_{ij}} e^{-i\left(q_{i}-q_{j}\right)}= 2e^{\frac{-i\alpha x}{x^{2}-1}}
\left[2-\left(\frac{\frac{1}{x}-X^{+}}{\frac{1}{x}-X^{-}}\right)^{2}\frac{X^{-}}{X^{+}}
-\left(\frac{x-X^{-}}{x-X^{+}}\right)^2\frac{X^{+}}{X^{-}}\right].\label{ps}
\end{equation}
For the big magnon, using the quasi-momenta
\begin{eqnarray}
q_{1}&=& -q_{10}= \frac{\alpha x}{x^{2}-1} \nonumber\\
q_{2}&=& -q_{9}= \frac{\alpha x}{x^{2}-1} \nonumber\\
q_{3}&=& -q_{8}= \frac{\alpha x}{x^{2}-1}-i \log\left(\frac{X^{+}}{X^{-}}\right)
+i \log\left(\frac{\frac{1}{x}-X^{+}}{\frac{1}{x}-X^{-}}\right)
-i \log\left(\frac{x-X^{+}}{x-X^{-}}\right)-p \nonumber\\
q_{4}&=& -q_{7}= \frac{\alpha x}{x^{2}-1}-i \log\left(\frac{X^{+}}{X^{-}}\right)
+i \log\left(\frac{\frac{1}{x}-X^{+}}{\frac{1}{x}-X^{-}}\right)
-i \log\left(\frac{x-X^{+}}{x-X^{-}}\right)-p\nonumber\\
q_{5}&=& -q_{6}= 0,\label{b}
\end{eqnarray}
we get
\begin{equation}
\sum_{ij} \left(-1\right)^{F_{ij}} e^{-i\left(q_{i}-q_{j}\right)}
= 4e^{\frac{-i\alpha x}{x^{2}-1}}\left[1-\frac{\frac{1}{x}-X^{+}}
{\frac{1}{x}-X^{-}}\frac{x-X^{-}}{x-X^{+}}\right].\label{big}
\end{equation}
The term $-p$ in (\ref{ps}) and (\ref{b}) are the twists of $\mathbb{RP}^{3}$ magnon and big magnon.

%%%%%%%%%%%%%%%%%%%%%%%%%%%%%%%%%%%%%

\section{$S$-matrix and L\"uscher formula\label{sec3}}
We use the Ahn-Nepomechie $S$-matrix of the \adsiv \,\ \cite{Ahn:2008aa} in the $F$-term L\"uscher formula
to compute the quantum corrections for the various dyonic GM states.

\subsection{$S$-matrix}
The ${\cal N}=6$ Chern-Simons theory contains two types of fundamental excitations, denoted by $A$ and $B$,
which belong to a fundamental representation of the centrally extended $su(2|2)$.
The $S$-matrices among these states are given by
\begin{eqnarray}
S^{AA}\left(p_{1},p_{2}\right) &=& S^{BB}\left(p_{1},p_{2}\right)
= S_{0}\left(p_{1},p_{2}\right) \hat S\left(p_{1},p_{2}\right) \cr
S^{AB}\left(p_{1},p_{2}\right) &=& S^{BA}\left(p_{1},p_{2}\right)
= \tilde S_{0}\left(p_{1},p_{2}\right) \hat S\left(p_{1},p_{2}\right),\nonumber
\end{eqnarray}
where $\hat S$ is the $su\left(2|2\right)$-invariant $S$-matrix \cite{Arutyunov:2006yd,Beisert:2005tm}.
Main feature of the $S$-matrix is encoded into the two scalar factors,
\begin{eqnarray}
S_{0}\left(p_{1},p_{2}\right) &=& \frac{1 - \frac{1}{x_{1}^{+} x_{2}^{-}}}{1 - \frac{1}{x_{1}^{-} x_{2}^{+}}}
\sigma\left(p_{1},p_{2}\right) \cr
\tilde S_{0}\left(p_{1},p_{2}\right) &=& \frac{x_{1}^{-} - x_{2}^{+}}{x_{1}^{+} - x_{2}^{-}}
\sigma\left(p_{1},p_{2}\right),\nonumber
\end{eqnarray}
where $\sigma\left(p_{1},p_{2}\right)$ is the BES dressing factor \cite{Beisert:2006ez}.
%%%%%%%%%%
These scalar factors have been checked classically in \cite{HatTan}.
%%%%%%%%%

Here we compute the leading one-loop correction using the $F$-term L\"uscher formula which is given by
\cite{Luscher:1985dn,Klassen:1990ub,Janik:2007wt}
\begin{equation}
\delta E_{F}= -\int \frac{dq}{2\pi}
\left[1- \frac{\varepsilon_{Q}'\left(p\right)}{\varepsilon_1'\left(q^{*}\right)}\right]
e^{-i q^{*} L}\sum_{b} \left(-1\right)^{F_{b}}
\left(S^{ba}_{ba}\left(q^{*},p\right)-1\right).\label{lus}
\end{equation}
The integrand of the formula consists of the kinematic and the $S$-matrix factors.

For the kinematical factor we consider first the ${\mathbb{CP}^{2}}$ small magnon whose
energy dispersion relation is given by
\begin{eqnarray}
\Delta - J/2 =\varepsilon_{Q}\left(p\right) = \sqrt{\frac{Q^{2}}{4}+16g^{2}\sin^{2}\frac{p}{2}}.\nonumber
\end{eqnarray}
It is convenient to introduce $X^{\pm}$ and $y^{\pm}$ variables defined by
\begin{eqnarray}
X^{+} - X^{-}+ \frac{1}{X^{+}} - \frac{1}{X^{-}}&=&\frac{i Q}{2g}\cr
y^{+} - y^{-} + \frac{1}{y^{+}} - \frac{1}{y^{-}}&=&\frac{i}{2g}\cr
\frac{X^{+}}{X^{-}} = e^{ip},\qquad\frac{y^{+}}{y^{-}} &=& e^{iq^{*}}.\label{star}
\end{eqnarray}
To compare the kinematic factor with algebraic curve results,
we introduce a variable $x$ defined by \cite{Bombardelli:2008qd,Gromov:2008ie}.
\begin{equation}
x+\frac{1}{x}\pm\frac{i}{4g}=y^{\pm}+\frac{1}{y^{\pm}},\nonumber
\end{equation}
which leads to
\begin{equation}
y^{\pm}= x \pm \frac{i x^2}{4g\left(x^2-1\right)}\nonumber
\end{equation}
at strong coupling limit.
The dispersion relation of virtual fundamental GM in this limit becomes
\begin{equation}
\varepsilon_1\left(q^{*}\right)=\frac{g}{i}\left(y^{+}-y^{-}-\frac{1}{y^{+}}+\frac{1}{y^{-}}\right)
= \frac{1}{2}\frac{x^{2}+1}{x^{2}-1}.\nonumber
\end{equation}
From (\ref{star}), one gets
\begin{equation}
e^{iq^{*}}\simeq 1+iq^{*} = 1+i\frac{x}{g\left(x^{2}-1\right)}.\nonumber
\end{equation}
Also from $q^{2}+\varepsilon'^{2}\left(q^{*}\right)=0$, $q$ can be written as
\begin{equation}
q=\frac{i}{2}\frac{x^{2}+1}{x^{2}-1}.\nonumber
\end{equation}
From these, we can get
\begin{eqnarray}
\varepsilon_Q'\left(p\right) &=& g \left(\frac{X^{+}+X^{-}}{X^{+} X^{-}+1}\right)\cr
\varepsilon_1'\left(q^{*}\right) &=& g \left(\frac{2x}{x^{2}+1}\right).\nonumber
\end{eqnarray}
Inserting these into the kinematic factor and changing the integration variable to $x$,
one can get the same kinematic factor as that of the algebraic curve computation,
\begin{equation}
\delta E^{F}_{\rm one-loop} = \int \frac{dx}{2\pi i} \partial_{x} \Omega\left(x\right)
e^{- \frac {ixJ}{2g\left(x^{2}-1\right)}}\ \sum_{b} \left(-1\right)^{F_{b}}
S^{b1_{Q}}_{b1_{Q}}\left(q^{*},p\right).\nonumber
\end{equation}
For other types of dyonic GMs, the results stand in the same way.

The $S$-matrix factor in the L\"uscher formula contains scatterings between virtual and
physical particles.
In the leading order, only fundamental particles contribute to the sum over virtual particles
$b=1,2,3,4$ of both $A$ and $B$ types.
The physical particles are dyonic GMs in $su(2)$ sector which are bound states of $Q$ number of
bosonic particles with $su(2|2)$ index $a=1_{Q}=(11\ldots 1)$.
Since the physical particles carry $su(2|2)$ index $1$, the relevant $S$-matrix elements are
\begin{eqnarray}
{\hat S}_{11}^{11}&=&a_{1} (p_{1},p_{2}),\quad {\hat S}_{21}^{21}=a_{1} (p_{1},p_{2})+a_{2} (p_{1},p_{2}),\quad
{\hat S}_{31}^{31}={\hat S}_{41}^{41}=a_{6} (p_{1},p_{2}),\cr
a_{1} (p_{1},p_{2})&=& \frac{x_{2}^{-} - x_{1}^{+}}{x_{2}^{+} - x_{1}^{-}} \frac{\eta_{1}
\eta_{2}}{\tilde{\eta}_{1} \tilde{\eta}_{2}} \cr
  a_{2} (p_{1},p_{2})&=& \frac{\left(x_{1}^{-} - x_{1}^{+}\right)\left(x_{2}^{-} - x_{2}^{+}\right)
  \left(x_{2}^{-} - x_{1}^{+}\right)}{\left(x_{1}^{-} - x_{2}^{+}\right)
  \left(x_{2}^{-}x_{1}^{-} - x_{2}^{+}x_{1}^{+}\right)}
  \frac{\eta_{1}\eta_{2}}{\tilde{\eta}_{1} \tilde{\eta}_{2}} \cr
  a_{6} (p_{1},p_{2})&=& \frac{x_{1}^{+} - x_{2}^{+}}{x_{1}^{-} - x_{2}^{+}}
  \frac{\eta_{2}}{\tilde{\eta}_{2}}.\nonumber
\end{eqnarray}
We choose the phase factors $\eta_i$ in the string frame \cite{Arutyunov:2006yd}
\begin{equation}
\frac{\eta_{1}}{\tilde{\eta}_{1}} = \sqrt{\frac{x_{2}^{+}}{x_{2}^{-}}},\,\
\frac{\eta_{2}}{\tilde{\eta}_{2}} = \sqrt{\frac{x_{1}^{-}}{x_{1}^{+}}}.\nonumber
\end{equation}

The relevant $S$-matrix elements for the dyonic magnons are
${S^{AA}}^{b1_{Q}}_{b1_{Q}}$ and ${S^{BA}}^{b1_{Q}}_{b1_{Q}}$ which are given by \cite{Ahn:2008wd}
\begin{eqnarray}
{S^{AA}}^{b1_{Q}}_{b1_{Q}}&=&\prod_{k=1}^{Q}
\left(\frac{1-\frac{1}{y^{+}x_{k}^{-}}}{1-\frac{1}{y^{-}x_{k}^{+}}}
\sigma_{BES}\left(y,x_{k}\right){\hat S}_{b1}^{b1}(y,x_{k})\right)\label{sma}\\
{S^{BA}}^{b1_{Q}}_{b1_{Q}}&=&\prod_{k=1}^{Q}
\left(\frac{y^{-}-x_{k}^{+}}{y^{+}-x_{k}^{-}}
\sigma_{BES}\left(y,x_{k}\right){\hat S}_{b1}^{b1}(y,x_{k})\right).\label{sma1}
\end{eqnarray}
From these, we obtain
\begin{eqnarray}
\sum_{b}\left(-1\right)^{F_{b}}{S^{AA}}^{b1_{Q}}_{b1_{Q}}&=&\sigma_{BES}
\left(y,X\right)\frac{\eta\left(X\right)}{\tilde{\eta}\left(X\right)}
\left(\frac{\eta\left(y\right)}{\tilde{\eta}\left(y\right)}\right)^{Q}
\left(s_{1}+s_{2}-s_{3}-s_{4}\right)S_{BDS}^{Q}\cr
\sum_{b}\left(-1\right)^{F_{b}}{S^{BA}}^{b1_{Q}}_{b1_{Q}}&=&\sigma_{BES}
\left(y,X\right)\frac{\eta\left(X\right)}{\tilde{\eta}\left(X\right)}
\left(\frac{\eta\left(y\right)}{\tilde{\eta}\left(y\right)}\right)^{Q}
\left(s_{1}+s_{2}-s_{3}-s_{4}\right).\nonumber
\end{eqnarray}
where we defined $x_{1}^{-}=X^{-}$, $x_{Q}^{+}=X^{+}$, and $x_{k}^{-}=x_{k-1}^{+}$.
The $s_b$s defined by
\begin{eqnarray}
s_{b}=\prod_{k=1}^{Q} \frac{{\hat S}_{b1}^{b1}(y,x_{k})}{{\hat S}_{11}^{11}(y,x_{k})}\nonumber
\end{eqnarray}
can be written as
\begin{eqnarray}
s_{1}&=&1 ,\qquad s_{2}=\left(\frac{y^{+}-X^{+}}{y^{+}-X^{-}}\right)
\left(\frac{1-\frac{1}{y^{-}X^{+}}}{1-\frac{1}{y^{-}X^{-}}}\right)\cr
s_{3}&=&s_{4}=\left(\frac{y^{+}-X^{+}}{y^{+}-X^{-}}\right)
\frac{\tilde{\eta}\left(X\right)}{\eta\left(X\right)}.
\label{sdef}
\end{eqnarray}
The dressing factor and $S_{BDS}$ of bound-state in this limit are given by
\cite{Hatsuda:2008gd,Gromov:2008ie}
\begin{eqnarray}
\sigma_{BES}\left(y,X\right)&=& \left(\frac{y-\frac{1}{X^{+}}}{y-\frac{1}{X^{-}}}\right)
e^{\frac{-ix}{2g\left(x^{2}-1\right)}\left(\Delta-J/2-Q\right)}\cr
S_{BDS}^{Q}&=& \frac{\left(y^{+}-X^{-}\right)\left(1-\frac{1}{y^{+}X^{-}}\right)}
{\left(y^{-}-X^{+}\right)\left(1-\frac{1}{y^{-}X^{+}}\right)}
\frac{\left(y^{-}-X^{-}\right)\left(1-\frac{1}{y^{-}X^{-}}\right)}
{\left(y^{+}-X^{+}\right)\left(1-\frac{1}{y^{+}X^{+}}\right)}.\label{SBDS}
\end{eqnarray}

\subsection{Small dyonic magnon ($\mathbb{CP}^{2}$)}
Dyonic small magnon is a bound-state of $A$-particles or $B$-particles.
Because the $S$-matrices are invariant under $A\leftrightarrow B$,
we can only consider a $A$-particle without loss of generality.
Then we can write the $S$-matrix elements for small dyonic magnon as below:
\begin{equation}
\sum_b \left(-1\right)^{F_{b}}\ S^{b1_{Q}}_{b1_{Q}}\left(q^{*},p\right) =
\sum_{b}\left(-1\right)^{F_{b}}\left({S^{AA}}^{b1_{Q}}_{b1_{Q}} + {S^{BA}}^{b1_{Q}}_{b1_{Q}}\right).
\end{equation}
We can approximate $y^{\pm}=x$ in the strong coupling limit for
the leading $F$-term integration and get
\begin{equation}
\frac{\eta\left(X\right)}{\tilde{\eta}\left(X\right)}= e^{\frac{ip}{2}},\qquad
\left(\frac{\eta\left(y\right)}{\tilde{\eta}\left(y\right)}\right)^{Q}
= e^{-i\frac{x Q}{2g\left(x^{2}-1\right)}}.
\end{equation}
These lead to the one-loop correction as follows:
\begin{eqnarray}
\delta\Delta_{F}&=&\int \frac{dx}{2\pi i} \partial_{x}\Omega\left(x\right)
e^{-i\Delta\frac{x}{2g\left(x^{2}-1\right)}}\left[
\left(\frac{x-\frac{1}{X^{+}}}{x-\frac{1}{X^{-}}}\right)e^{\frac{ip}{2}}
+\left(\frac{x-X^{-}}{x-X^{+}}\right)e^{\frac{ip}{2}}\right.\nonumber\\
&&+\left(\frac{x-\frac{1}{X^{-}}}{x-\frac{1}{X^{+}}}\right)\left(\frac{x-X^{-}}{x-X^{+}}\right)^{2}
e^{\frac{ip}{2}}
+\left(\frac{x-\frac{1}{X^{+}}}{x-\frac{1}{X^{-}}}\right)^{2}\left(\frac{x-X^{+}}{x-X^{-}}\right)
e^{\frac{ip}{2}}\nonumber\\
&&\left.- 2\left(\frac{x-\frac{1}{X^{+}}}{x-\frac{1}{X^{-}}}\right)
\left(\frac{x-X^{+}}{x-X^{-}}\right)- 2\left(\frac{x-\frac{1}{X^{-}}}{x-\frac{1}{X^{+}}}\right)
\left(\frac{x-X^{-}}{x-X^{+}}\right)\right].
\end{eqnarray}
It can be easily seen that the expression in bracket with the exponential factor in front
is exactly the same as the result in (\ref{small}) in algebraic curve for small dyonic magnon.

\subsection{Pair of small dyonic magnon ($\mathbb{RP}^{3}$)}

In case of $\mathbb{RP}^{3}$ dyonic magnon, we need to consider a pair of
$A$ and $B$ type dyonic GMs in the physical state with the same momentum $p_{1}=p_{2}=p$
\cite{Ahn:2008wd}.
The $F$-term L\"uscher formula becomes \cite{Bombardelli:2008qd}
\begin{eqnarray}
\sum_b \left(-1\right)^{F_{b}}\ S^{b1_{Q}}_{b1_{Q}}\left(q^{*},p\right)
&=& \sum_{b}\left(-1\right)^{F_{b}}\left({S^{AA}}^{b1_{Q}}_{b1_{Q}}{S^{AB}}^{b1_{Q}}_{b1_{Q}}
+ {S^{BA}}^{b1_{Q}}_{b1_{Q}}{S^{BB}}^{b1_{Q}}_{b1_{Q}}\right)\nonumber\\
&=&2\sum_{b} \left(-1\right)^{F_{b}} {S^{AA}}^{b1_{Q}}_{b1_{Q}}{S^{BA}}^{b1_{Q}}_{b1_{Q}}.
\end{eqnarray}
Using the $S$-matrix elements in (\ref{sma}) and (\ref{sma1}), we obtain
\begin{equation}
\delta\Delta_{F}=\int \frac{dx}{2\pi i} \partial_{x}\Omega\left(x\right)
e^{-i\Delta\frac{x}{2g\left(x^{2}-1\right)}}
\left[4-2\left(\frac{x-\frac{1}{X^{+}}}{x-\frac{1}{X^{-}}}\right)^{2}
e^{ip}-2\left(\frac{x-X^{-}}{x-X^{+}}\right)^{2}e^{ip}\right].\nonumber
\end{equation}
By comparing this with (\ref{ps}) of algebraic curve, one can confirm that
the energy shift from both methods agree each other.

\subsection{Big magnon}

Particle interpretation of big magnon is not clear until now.
We propose that big magnon may be superposition of a small magnon and an ``anti-small magnon''.
The anti-small magnon has the same momentum as the usual small magnon but the quantum number $-Q$.
Then, we can compute $S$-matrix elements as below:
\begin{equation}
\sum_b \left(-1\right)^{F_{b}}\ S^{ba}_{ba}\left(q^{*},p\right) =
\sum_{b}\left(-1\right)^{F_{b}}\left({S^{AA}}^{b1_{Q}}_{b1_{Q}}{S^{AA'}}^{b1_{-Q}}_{b1_{-Q}}
+ {S^{BA}}^{b1_{Q}}_{b1_{Q}}{S^{BA'}}^{b1_{-Q}}_{b1_{-Q}}\right).\nonumber
\end{equation}
From these $S$-matrix elements, we have
\begin{eqnarray}
&&\sum \left(-1\right)^{F_{b}}\left(S^{ba}_{ba}\left(q^{*},p\right)\right)
=2\prod_{k=1}^{Q}\left[\left(\frac{1-\frac{1}{y^{+}X_{k}^{-}}}{1-\frac{1}{y^{-}X_{k}^{+}}}
\right)\left(\frac{y^{+}-X_{k}^{-}}{y^{-}-X_{k}^{+}}\right)
\left(\frac{1-\frac{X_{k}^{+}}{y^{+}}}{1-\frac{X_{k}^{-}}{y^{-}}}\right)
\left(\frac{y^{+}-\frac{1}{X_{k}^{+}}}{y^{-}-\frac{1}{X_{k}^{-}}}\right)\right]\cr
&&\times\left(\frac{y-\frac{1}{X_{Q}^{+}}}{y-\frac{1}{X_{Q}^{-}}}\right)
\left(\frac{y-X_{Q}^{-}}{y-X_{Q}^{+}}\right)\left(\frac{\eta\left(X^{Q}\right)}{\tilde{\eta}
\left(X^{Q}\right)}\right)^{2}\
e^{-i\left(\Delta-J\right)\frac{x}{2g\left(x^{2}-1\right)}}\
\left(1+s_{2}s'_{2}-2s_{3}s'_{3}\right).\nonumber
\end{eqnarray}
The primed quantities, $A'$,$s'_{2}$ and $s'_{3}$, are related to those unprimed by
$X^{\pm}\to\frac{1}{X^{\mp}}$ which is equivalent to changing $Q\to -Q$.
From the $S$-matrix element, the one-loop correction of big magnon is given by
\begin{equation}
\delta\Delta_{F} =\int \frac{dx}{2\pi i} \partial_{x}\Omega\left(x\right)
e^{-i\Delta\frac{x}{2g\left(x^{2}-1\right)}}
4\left[1-\left(\frac{x-\frac{1}{X^{+}}}{x-\frac{1}{X^{-}}}\right)
\left(\frac{x-X^{-}}{x-X^{+}}\right)e^{ip}\right].\nonumber
\end{equation}
This result matches exactly with that of the algebraic curve (\ref{big}).
The one-loop corrections for pair of small magnon and big magnon are exactly the same in
the non-dyonic limit.
While the physical meaning of the anti-magnon state is not clear, the one-loop correction analysis
suggests how to express the big magnon state in terms of the on-shell particles.

\section{Multi dyonic magnons\label{sec4}}
As authors of \cite{Lukowski:2008eq,Minahan:2008re,Sax:2008in} first explain, in algebraic curve
technology, the case of $(N,M)$-type magnon is realized as a sum of $N$ resolvents of $u$ and $M$ resolvents of $v$ type.
The meaning of the $r$-type resolvent is not clear in the $S$-matrix interpretation.
So, we set the $r$-type resolvent to zero and consider only $N$-number of $u$-type excitations and
$M$-number of $v$-type excitations.
These correspond to general excitations of $N$-number of $A$-type dyonic magnons
and $M$-number of $B$-type dyonic magnons in spin chain.
$\mathbb{CP}^2$ and $\mathbb{RP}^3$ magnons are special cases of this general multi-magnon state.
To compute one-loop finite size effects for such configurations of magnons,
we use the following quasi-momenta ansatz:
\begin{eqnarray}
q_{1}&=& -q_{10}= \frac{\alpha x}{x^{2}-1} \nonumber\\
q_{2}&=& -q_{9}= \frac{\alpha x}{x^{2}-1} \nonumber\\
q_{3}&=& -q_{8}= \frac{\alpha x}{x^{2}-1}+ \sum_{k=1}^{N}\left(G_{u}^{k}\left(0\right)-
G_{u}^{k}\left(\frac{1}{x}\right)\right)+\sum_{k=N+1}^{N+M}\left(G_{v}^{k}\left(0\right)-
G_{v}^{k}\left(\frac{1}{x}\right)\right)+\sum_{i=1}^{N+M}\tau_{i} \nonumber\\
q_{4}&=& -q_{7}= \frac{\alpha x}{x^{2}-1}+ \sum_{k=1}^{N}G_{u}^{k}\left(x\right)
+\sum_{k=N+1}^{N+M}G_{v}^{k}\left(x\right)+\sum_{i=1}^{N+M}\tau_{i} \nonumber\\
q_{5}&=& -q_{6}= \sum_{k=1}^{N}\left(G_{u}^{k}\left(x\right)-G_{u}^{k}\left(0\right)
+ G_{u}^{k}\left(\frac{1}{x}\right)\right)+\sum_{k=N+1}^{N+M}\left(-G_{v}^{k}\left(x\right)
+G_{v}^{k}\left(0\right)- G_{v}^{k}\left(\frac{1}{x}\right)\right).\nonumber
\end{eqnarray}
Here, $G_{u}^{k}\left(x\right)$ and $G_{v}^{k}\left(x\right)$ are all the same with
$G_{\rm{magnon}}=-i \log\left(\frac{x-X^{+}_{k}}{x-X^{-}_{k}}\right)$ and
$\tau_{i}=- \frac{p_{i}}{2}$.

The problem here is that unknown functions in the fluctuations of quasi-momenta
can not be completely determined by the constraint equations coming from asymptotic limit.
So we get some undetermined functions $\alpha_{l}$ which satisfy
$\sum_{l}\alpha_{l}=1$ where the index $l$ runs from $1$ to $N+M$.
This feature of the fluctuation frequencies for multi-particle states is noticed first in
\cite{Hatsuda:2008na} for ${\cal N}=4$ SYM.
In \adsiv, off-shell frequencies are as below.
\begin{eqnarray}
\Omega^{\rm light}_{ij}\left(x\right)&=&\frac{1}{x^{2}-1}
\left(1- \sum_{l} \alpha_{l}\left(x \frac{X^{+}_{l}+X^{-}_{l}}{X^{+}_{l}X^{-}_{l}+1}\right)\right)=\Omega_{\rm multi}\left(x\right)
\cr
\Omega^{\rm heavy}_{ij}\left(x\right)&=&\frac{2}{x^{2}-1}
\left(1- \sum_{l} \alpha_{l}\left(x \frac{X^{+}_{l}+X^{-}_{l}}{X^{+}_{l}X^{-}_{l}+1}\right)\right).\nonumber
\end{eqnarray}
We obtain the integral representation of the one-loop effect as before by using
$\Omega_{\rm multi}\left(x\right)$ and quasi-momenta for multi-magnons configurations.
\begin{eqnarray}
\sum_{ij} \left(-1\right)^{F_{ij}} e^{-i\left(q_{i}-q_{j}\right)}
&=&\left(\prod^{N}_{i=1}\left(\frac{x-\frac{1}{X^{-}_{i}}}{x-\frac{1}{X^{+}_{i}}}\right)^{2}
\left(\frac{x-X^{-}_{i}}{x-X^{+}_{i}}\right)^{2}
+\prod^{N+M}_{j=N+1}\left(\frac{x-\frac{1}{X^{-}_{j}}}{x-\frac{1}{X^{+}_{j}}}\right)^{2}
\left(\frac{x-X^{-}_{j}}{x-X^{+}_{j}}\right)^{2}\right)\cr
&\times&\left(1+\prod^{M+N}_{i=1}\left(\frac{x-\frac{1}{X^{+}_{i}}}{x-\frac{1}{X^{-}_{i}}}
\right)\left(\frac{x-X^{+}_{i}}{x-X^{-}_{i}}\right)
-2\prod^{N+M}_{j=1}\left(\frac{x-X^{+}_{j}}{x-X^{-}_{j}}\right)
\sqrt{\frac{X^{-}_{i}}{X^{+}_{i}}}\right)\cr
&\times&\prod^{N+M}_{i=1}\left(\frac{x-\frac{1}{X^{+}_{i}}}{x-\frac{1}{X^{-}_{i}}}\right)
\sqrt{\frac{X^{+}_{i}}{X^{-}_{i}}}.\label{mlt}
\end{eqnarray}
Here $\left(i,j\right)$ pairs include only light modes.
To compare this with multi-particle L\"uscher formula, we need to compute the $S$-matrix
factor which includes scatterings between a virtual fundamental magnon and
multi dyonic magnons.
The $S$-matrix factor is given by
\begin{eqnarray}
&&\sum\left(-1\right)^{F_{b}}S_{\rm multi}=\sigma_{BES}\left(y,X_{1}^{Q_{1}}\right)
\cdots\sigma_{BES}\left(y,X_{N+M}^{Q_{N+M}}\right)\frac{\eta\left(X_{1}^{Q_{1}}\right)}
{\tilde{\eta}\left(X_{1}^{Q_{1}}\right)}\cdots\frac{\eta\left(X_{N+M}^{Q_{N+M}}\right)}
{\tilde{\eta}\left(X_{N+M}^{Q_{N+M}}\right)}\cr
&\times&\left(\prod^{N}_{i=1}S_{BDS}\left(y,X_{i}^{Q_{i}}\right)+\prod^{N+M}_{i=N+1}S_{BDS}
\left(y,X_{i}^{Q_{i}}\right)\right)\
\left(\frac{\eta\left(y\right)}{\tilde{\eta}\left(y\right)}\right)^
{\sum Q_{i}}\sum_{b}(-1)^{F_b}\prod_{i=1}^{N+M} s_b(p_i).\nonumber
\end{eqnarray}
Using $s_b(p)$ defined in (\ref{sdef}) and $S_{BDS}$ in (\ref{SBDS}), we can obtain
the final expression
\begin{eqnarray}
\delta\Delta_{F}^{\rm multi}&=&\int \frac{dx}{2\pi i} \partial_{x}\Omega_{multi}\left(x\right)
e^{-i \Delta_{total}\frac{x}{2g\left(x^{2}-1\right)}}
\times\prod^{N+M}_{i=1}\left(\frac{x-\frac{1}{X^{+}_{i}}}{x-\frac{1}{X^{-}_{i}}}
\sqrt{\frac{X^{+}_{i}}{X^{-}_{i}}}\right)\cr
&\times&\left(\prod^{N}_{i=1}\left(\frac{x-X^{-}_{i}}{x-X^{+}_{i}}\right)^{2}
\left(\frac{x-\frac{1}{X^{-}_{i}}}{x-\frac{1}{X^{+}_{i}}}\right)^{2}+\prod^{N+M}_{j=N+1}
\left(\frac{x-X^{-}_{j}}{x-X^{+}_{j}}\right)^{2}\left(\frac{x-\frac{1}{X^{-}_{j}}}{x-\frac{1}
{X^{+}_{j}}}\right)^{2}\right)\cr
&\times&\left(1+\prod^{N+M}_{k=1}\left(\frac{x-X^{+}_{k}}{x-X^{-}_{k}}\right)
\left(\frac{x-\frac{1}{X^{+}_{k}}}{x-\frac{1}{X^{-}_{k}}}\right)
-2\prod^{N+M}_{l=1}\left(\frac{x-X^{+}_{l}}{x-X^{-}_{l}}\right)
\sqrt{\frac{X^{-}_{l}}{X^{+}_{l}}}\right).\nonumber
\end{eqnarray}
Here we have used multi-particle L\"uscher formula \cite{Hatsuda:2008na}.
It is straightforward to check that this result matches with the algebraic curve result
(\ref{mlt}).

\section{Conclusions\label{sec5}}
In this paper we have computed the quantum finite-size effects for $\mathbb{CP}^2$,
$\mathbb{RP}^3$, and big magnons and general combination of these magnons.
We have computed the off-shell frequencies of magnons
from the fluctuations of quasi-momenta and summed over all the physical polarizations.
We have shown that these results agree with the $F$-term L\"uscher formula based on
the \adsiv\ $S$-matrix proposed in \cite{Ahn:2008aa}.
This provides a stringent check for the $S$-matrix up to the one-loop order
as well as the off-shell algebraic curve method.

In the computations, we have noticed that the unstable heavy modes should be
suppressed in the $AdS_{4}\times\mathbb{CP}^{3}$ algebraic curve.
We also provide an on-shell particle interpretation of the big magnon, which was found in
the context of the algebraic curve \cite{Shenderovich:2008bs} and the dressing methods
\cite{Hollowood:2009tw,Kalousios:2009mp,Suzuki:2009sc}, as a bound-state of
two small dyonic magnons with $u(1)$ charges $Q$ and $-Q$, respectively.
Our result shows that the L\"uscher $F$-term based on this particle interpretation
correctly reproduces the one-loop energy shift from the algebraic curve method.
It is not clear at this moment if this is just mathematical coincidence or has a deeper
physical meaning.
In the \adsv\ context, it has been argued that
a dyonic magnon state with $-Q$ is an anti-magnon state which is obtained by
reversing time direction \cite{DHM}.
It will be interesting if this argument also applies to \adsiv.

The undetermined functions $\alpha_{l}$ arising in the algebraic curve method for multi-magnon states
should be determined in the $S$-matrix approach.
For example, the multi-particle L\"uscher formula in \cite{Bajnok:2008bm}
doesn't include any such functions.
In fact, these functions are related with the variation of Bethe-Yang equations.
Therefore, we can compute $\alpha_{l}$ from Bethe-Yang equations.
In the case of the strong coupling limit, terms that contribute to $\alpha_{l}$ vanish in the
$F$-term integration for multi-magnons by using saddle point method.
However, it becomes important to determine these $\alpha_{l}$ at weak coupling regime
such as in the five-loop Konishi computation \cite{Bajnok:2009vm}.

{\bf Note added:}

While typing this article, we have found a paper \cite{Abbott:2010yb} whose results on the off-shell
algebraic curves overlap partially with ours.

\section*{Acknowledgements}
We thank D. Bombardelli and I. Aniceto for informing us their result \cite{Abbott:2010yb} and valuable comments.
This work was supported in part by KRF-2007-313- C00150, WCU Grant No. R32-2008-000-101300 (C. A. and M. K.),
and by the National Research Foundation of Korea (NRF) grant funded by the Korea government (MEST)
through the Center for Quantum Spacetime (CQUeST) of Sogang University with grant number 2005-0049409
(M. K. and B.-H. L.).

\end{document}